\documentclass[twocolumn]{aa}
      \def\folio{\ifnum\pageno=1\nopagenumbers\else\number\pageno\fi}

\def\lax    {\ifmmode{_<\atop^{\sim}}\else{${_<\atop^{\sim}}$}\fi}
\def\gax    {\ifmmode{_>\atop^{\sim}}\else{${_>\atop^{\sim}}$}\fi}
\newbox\grsign      \setbox\grsign=\hbox{$>$}
\newdimen\grdimen   \grdimen=\ht\grsign
\newbox\simgreatbox \setbox\simgreatbox=\hbox{\raise.5ex\hbox{$>$}\llap
                        {\lower.5ex\hbox{$\sim$}}}\ht1=\grdimen\dp1=0pt
\newbox\simlessbox  \setbox\simlessbox =\hbox{\raise.5ex\hbox{$<$}\llap
                        {\lower.5ex\hbox{$\sim$}}}\ht2=\grdimen\dp2=0pt

\newbox\grsign \setbox\grsign=\hbox{$>$} \newdimen\grdimen \grdimen=\ht\grsign
\newbox\laxbox \newbox\gaxbox
\setbox\gaxbox=\hbox{\raise.5ex\hbox{$>$}\llap
     {\lower.5ex\hbox{$\sim$}}}\ht1=\grdimen\dp1=0pt
\setbox\laxbox=\hbox{\raise.5ex\hbox{$<$}\llap
     {\lower.5ex\hbox{$\sim$}}}\ht2=\grdimen\dp2=0pt
\def\gax{\mathrel{\copy\gaxbox}}
\def\lax{\mathrel{\copy\laxbox}}
\def\boxit#1    {\vbox{\hrule\hbox{\vrule\kern3pt
                  \vbox{\kern3pt#1\kern3pt}\kern3pt\vrule}\hrule}}
\def\h      {\ifmmode{^{\rm h}}\else{$^{\rm h}$}\fi}
\def\m      {\ifmmode{^{\rm m}}\else{$^{\rm m}$}\fi}
\def\s      {\ifmmode{^{\rm s}}\else{$^{\rm s}$}\fi}
\def\decas    {\ifmmode{{\rlap.}{''}}\else{${\rlap.}{''}$}\fi}
\def\mum     {\ifmmode{\mu{\rm m}}\else{$\mu{\rm m}$}\fi}
\def\s      {\ifmmode{^{\rm s}}\else{$^{\rm s}$}\fi}
\def\decdeg {\rlap . {}^\circ}     
\def\deg      {\ifmmode{^{\circ}}\else{$^{\circ}$}\fi}
\def\as     {\ifmmode {\rlap.}$\,$''$\,$\! \else ${\rlap.}$\,$''$\,$\!$\fi}
\def\decsec  {\ifmmode {\rlap.}$\,$^{s}$\,$\! \else ${\rlap.}$\,$^{s}$\,$\!$\fi}\def\decs  {\ifmmode {\rlap.}$\,$^{s}$\,$\! \else ${\rlap.}$\,$^{s}$\,$\!$\fi}

\def\kms    {\ifmmode{{\rm km~s}^{-1}}\else{km~s$^{-1}$}\fi}

\def\Lsun   {$L_{\odot}$}

\def\Mspy   {\ifmmode {M_{\odot} {\rm yr}^{-1}} \else $M_{\odot}$~yr$^{-1}$\fi}
\def\Mdot   {\ifmmode {\dot M} \else $\dot M$\fi}
\def\mhd    {\ifmmode {n_{{\rm H}_2}} \else $n_{{\rm H}_2}$\fi}
\def\mhcd   {\ifmmode {N_{{\rm H}_2}} \else $N_{{\rm H}_2}$\fi}

\def\El      {\ifmmode{E_{\ell}}\else{$E_{\ell}$}\fi}
\def\beam    {\ifmmode{\theta_{\rm B}}\else{$\theta_{\rm B}$}\fi}
\def\mjyb   {\ifmmode {{\rm mJy~beam}^{-1}} \else{mJy~beam$^{-1}$}\fi}
\def\mujyb   {\ifmmode {\mu{\rm Jy~beam}^{-1}} \else{$\mu$Jy~beam$^{-1}$}\fi}
\def\Trot   {\ifmmode{T_{\rm rot}}\else$T_{\rm rot}$\fi}

\def\Teff   {\ifmmode{T_{\rm eff}}\else$T_{\rm eff}$\fi}

\def\ITRS   {\ifmmode{\smallint {\rm T}_{R}^{*}dv}\else{$\smallint
{\rm T}_{R}^{*}dv$}\fi}
\def\ITRS   {\ifmmode{\smallint {\rm T}_{R}^{*}dv}\else{$\smallint
{\rm T}_{R}^{*}dv$}\fi}
\def\ITAS   {\ifmmode{\smallint {\rm T}_{A}^{*}dv}\else{$\smallint
{\rm T}_{A}^{*}dv$}\fi}

\usepackage{graphicx}
\usepackage{natbib}
\usepackage{ulem}
      \def\new#1 {{\bf #1 }}
      \def\cut#1 {\sout{#1} }

\def\pmin {\phantom{$-$}}

\def\ra {$\rightarrow$}

\begin{document}

\title{Radio continuum monitoring of the extreme carbon star IRC+10216}
\author{K. M. Menten
\inst{1}
\and
M. J. Reid
\inst{2}
\and
E. Kr\"ugel
\inst{1}
\and
M. J. Claussen
\inst{3}
\and
R. Sahai
\inst{4}
}

\offprints{K. M. Menten}

\institute{Max-Planck-Institut f\"ur Radioastronomie,
Auf dem H\"ugel 69, D-53121 Bonn, Germany
\email{kmenten,p309ekr@mpifr-bonn.mpg.de}
\and
Harvard-Smithsonian Center for Astrophysics,
60 Garden Street, Cambridge MA 02138, USA
\email{reid@cfa.harvard.edu}
\and
National Radio Astronomy Obsrvatory,
Array Operations Center, P.O. Box O,
Socorro,
NM 87801, USA
\email{mclausse@nrao.edu}
\and
Jet Propulsion Laboratory,
MS 183-900,
4800 Oak Grove Drive,
Pasadena, CA 91109, USA
\email{raghvendra.sahai@jpl.nasa.gov}}

\date{Received / Accepted}
\titlerunning{Radio continuum observations of IRC+10216}
\authorrunning{Menten et al.}

\abstract{We describe Very Large Array observations of the extreme carbon star
IRC+10216 at 8.4, 14.9, and 22.5 GHz made over a two year period.  We find
possible variability correlated with the infrared phase and a cm- to
sub-millimeter wavelength spectral index very close to 2.
The variability, observed flux densities, and upper limit on the size
are consistent with the emission arising from the stellar photosphere or
a slightly larger radio photosphere.

\keywords{Stars: carbon --- Stars: individual: IRC+10216 --- Radio continuum: stars}}

\maketitle

\section{\label{intro}Introduction}

The ``extreme'' carbon star IRC+10216 (CW Leonis) was
discovered in the early days of modern infrared astronomy \citep{Becklin1969}.
The star is one of the brightest near- and mid-infrared sources
(LeBertre 1987 and references therein).
IRC+10216 has an extremely rich molecular spectrum arising from
a dense envelope created by a powerful mass outflow
\citep{Cernicharo2000}.  These attributes, together with its proximity
(distance of $\approx130$ pc) and its prodigious mass-loss rate (1.5
-- $3.3\times10^{-5}$ \Mspy), make IRC+10216 a ``keystone'' object for
which many aspects of (carbon-star) asymptotic giant branch (AGB) evolution
can be studied in detail. IRC+10216's properties are summarized in Table
\ref{hip}.

IRC+10216 has been detected at wavelengths of 1.5 and 2 cm
(\citet{Sahai1989}, \citet{Drake1991}).  From Very Large Array (VLA) observations it
should be possible to measure its size, position (which is extremely difficult
at other wavelengths owing to absorption and scattering), proper motion, and
the spectral index of its radio emission. Very Large Array data can strongly constrain
our information on the star's nature.

This paper has the following structure: In \S\ref{vlaobs}, we describe
six-epoch, three-wavelength VLA observations. These, and an additional
dataset (published by Drake et al. 1991), were retrieved from the VLA archive.
The properties of the continuum emission are discussed in \S\ref{properties}.
In particular, we address the variability, spectral index, absolute
position, and source morphology.
In \S\ref{nature} we synthesize these results to reach
conclusions on the nature of the radio emission.


\begin{table*}[tb]
\begin{center}
 \caption{\label{hip}IRC+10216  -- Fundamental Parameters}
 \begin{tabular}{lll}
 \hline \hline
Name & IRC+10216  & Reference\\
\noalign{\smallskip}
 \hline
 \noalign{\smallskip}
Aliases                               & CW Leonis, RAFGL 1381 & SIMBAD$^{a}$\\
                                      &  IRAS 09452+1330 \\
$\alpha_{{\rm J}2000}$                &\pmin$09\h47\m57\decs392^{b}$\\
$\delta_{{\rm J}2000}$                &   $+13\deg16'43\decas63^{b}$\\
Spectral type                         & C               & \\
Effective temperature, $T_{\rm eff}$      & 2000 K$^{c}$ & \citet{Groenewegen1997}  \\
Luminosity, $L_\star$                 &  $8.2~10^5~[D(kpc)]^2$ \Lsun$^{d}$  & \citet{Groenewegen1997} \\
Mass loss rate, $\Mdot$                & 1.5 -- $3.3\times10^{-5}$ \Mspy $^{e}$ & \citet{Crosas1997},\\
                                      && \citet{Groenewegen1998}\\
Distance, $D^{d}$                         & 110 -- 150 pc &\citet{Crosas1997},\\
				      && \citet{Groenewegen1998}\\
Expansion velocity                    & 16~\kms &\citet{Loup1993}\\
LSR velocity,  $v_{\rm LSR}$          & $-22\pm1.5$ \kms & \citet{Loup1993}\\
Heliocentric velocity, $v_{\rm hel}$  & $-15$ \kms         \\
 \noalign{\smallskip}
 \hline
 \noalign{\smallskip}
 \end{tabular}
\end{center}
$^{a}$http://simbad.u-strasbg.fr/
$^{b}$Position derived from 1993 January 26 VLA data. The estimated uncertainty
is 25 milli arcseconds (see \S\ref{astrometry}).
$^{c}$\citet{Groenewegen1997} calculates a range of models resulting
from a range of values for this parameter;
we adopt 2000~K, which is used in most of his models.
$^{d}$Assuming  $T_{\rm eff} = 2000$ K.
$^{e}$The listed mass loss rate and distance estimates represent  the range of values derived
from modeling of multi-line CO observations described in the referenced studies. \end{table*}

\section{\label{vlaobs} VLA Observations and data reduction}

\subsection{1991 -- 1993 observations }

The data reported here were taken by two of us (RS \&\ MC)
with the NRAO Very Large Array (VLA)\footnote{The Natinal Radio Astronomy
Observatory (NRAO) is operated by Associated Universities Inc., under a
collaborative agreement with the U.S. National Science Foundation.}.
We analyzed observations made between 1991 June 1 and 1993 June 9,
spaced typically by 3 to 4 months.
Observations were made at 8.4, 14.9, and 22.5 GHz (X-, U-, and K-band,
respectively)\footnote{To avoid confusion, throughout the paper radio
spectral bands are referred to in roman font and infrared bands
in \textit{italics}.}.  Details are given in Table \ref{runs}.
A typical observing run consisted two $\sim 20$ -- 28 minute long X- and
U-band observations, preceded and followed by
observations of the calibrator 0953$+$254. Because of the
generally short coherence times at the highest frequency, the K-band
scans were shorter ($\approx 15$ minute duration) and more
scans were made in this band.  At the end of each
run, we observed the absolute flux density calibrator 3C~286.

The flux density of 0953$+$254 was determined by comparing its visibility
amplitude with that of 3C~286.
The elevations of 3C~286 were comparable to that of 0953$+$254, minimizing
systematic amplitude calibration errors due to elevation-dependent antenna
gain.  By comparing the variations of the flux densities for several
scans of 0953$+$254, we estimate our flux density scale is
accurate to within 10\%\ at X-band, 15\%\ at U-band, and 20\% at K-band.
Data from one epoch (1992 October 2) suffered serious instrumental
difficulties and were discarded.
Table \ref{runs} contains the measured flux densities of 0953$+$254.
Complex gain correction factors were interpolated from the amplitude and
phase of 0953$+$254 to the times of the IRC+10216 data.


We made maps using the AIPS task IMAGR and
fit elliptical Gaussian models to the images.  The integrated
flux densities are listed in Table \ref{runs}.
Because IRC+10216 is weak, self calibration was not possible, and
the ``fast-switching'' technique, routinely employed now, had not been developed
at the time of our observations.  Nevertheless, during one epoch (1993 January
26) when the array was in the A-configuration, superb phase coherence persisted
throughout the observations. We use those data to establish the
source position (in \S\ref{astrometry}).

\subsection{\label{Other}Other VLA data}

VLA A-configuration observations of IRC+10216 have been reported by
\citet{Drake1991}.  We re-analyzed their 1987 June 2 and 3 U-band data
to compare with our U-band data.
Fitting an elliptical Gaussian using the AIPS task JMFIT, we
find the source to be nearly unresolved, with upper limits on the
deconvolved size of 95 milliarcesconds (mas) and a peak brightness and
integrated flux density of $1.29\pm0.15$ \mjyb\ and $1.40\pm0.27$ mJy,
respectively.
The latter value can be compared with our observed range of integrated
U-band flux densities of 1.77 to 2.67 mJy (see Table \ref{runs}).
Since we used identical calibration procedures for both datasets and
the same {\it absolute} flux calibrator, this discrepancy is
probably real.  Since the 1987 data employed a secondary gain
calibrator nearer to IRC+10216 than used for our data, one would expect
that the 1987 data should yield a greater flux density than the 1993 data,
if random atmospheric phase fluctuations degrade the imaging.

In order to estimate of the \textit{absolute} position accuracy, we made images
from the the June 2 and 3 data separately.
The positions of the fitted Gaussians differed by 18 mas in right ascension and
10 mas in declination, testifying to the excellent phase coherence during these
observations.  Assuming that large-scale atmospheric effects, which
usually limit astrometric accuracy, were independent for these two days,
this indicates an uncertainty of $\approx20$ mas in each coordinate.

\begin{table*}[!t]
\caption{\label{runs}Log of IRC+10216 multiband VLA observations}
\begin{center}
\begin{tabular}[!t]{lllllllll}
JD~2440000+& Civil Date &Array &\multicolumn{3}{l}{$S$(B0953$+$254)(Jy)} &\multicolumn{3}{l}{$S$(IRC+10216)(mJy)}  \\
         &            &      &X-band & U-band & K-band &X-band & U-band & K-band \\
\noalign{\smallskip}
\hline
8410 &  1991 June 01     & D(\ra A)& 1.80(0.01)& 1.90(0.05) &2.27(0.07) & 0.92(0.07) & 2.67(0.39) & 7.10(0.68)\\   
8524  & 1991 September 24& A\ra B  & 1.80(0.08)& 2.02(0.13) &2.40(0.13) & 0.48(0.07) & 2.11(0.55) & 3.87(1.20)\\   
8652  & 1992 February 09 & CnB     & 1.56(0.01)& 1.58(0.01) &1.51(0.01) & 0.82(0.04) & 2.23(0.17) & 4.24(0.33)\\   
8772  & 1992 May 30      & C\ra D  & 1.54(0.01)& 1.52(0.02) &1.33(0.03) & 0.86(0.07) & 2.42(0.20) & 3.18(0.67)\\   
9014  & 1993 January 26  & A       & 1.58(0.03)& 1.55(0.04) &1.60(0.05) & 0.60(0.04) & 2.00(0.31) & 5.33(1.03)\\   
9143  & 1993 June 09     & CnB     & 1.72(0.01)& 1.69(0.05) &1.77(0.24) & 0.48(0.04) & 1.77(0.21) & 3.86(0.60)\\   
\hline
Average&                 & --      & --        & --         & --        & 0.67(0.03) & 2.14(0.18) & 4.20(0.46)\\
\noalign{\smallskip}
\hline
\noalign{\smallskip}
 \end{tabular}
\end{center}

The first six lines present the results of our VLA measurements epoch by epoch.
The first and second columns give the Julian and civil dates, respectively.
The third column
gives the array configuration. Two letters connected by an arrow
denote epochs when the array was in transition from one to another
configuration; CnB denotes a mixed C- and D-configuration.
The forth to sixth columns give the
bootstrapped flux density of the phase calibrator B0953$+$25 at X-,
U-, and K-band respectively, while the seventh to ninth columns give
the same for IRC+10216. The quoted uncertainties are formal errors
from JMFIT.
We estimate additional systematic errors of 10, 15, and 20\%\
for X-, U-, and K-band data, respectively.
Based on the formulae given by \citet{Baars1977}, the absolute
flux calibrator 3C~286 was assumed to have flux densities of 5.21,
3.46, and 2.52 Jy at X-, U-, and K-band.  For each band, data from two 50 MHz
wide channels were taken in right- and left-hand circular polarization.
The center frequencies were 8.439, 14.940, and 22.460 GHz for the X-,
U-, and K-band data.  For the first epoch (1991 June 1, JD 2448410) only
visibilities within a $u,v$-range of 50 k$\lambda$, corresponding to D-array
antenna separations, were considered.  Data from one more epoch (1992 October 2,
JD 2448897) had to be discarded due to instrumental problems.  The bottom line
gives weighted averages, with individual-epoch uncertainties estimated by
adding quadratically the formal uncertainties from JMFIT
with estimated absolute calibration uncertainties.
\end{table*}

\section{\label{properties}Properties of the radio emission}

\subsection{\label{variability}Variability}

\begin{figure}[h]
\begin{center}
\includegraphics[width=8cm,angle=0]{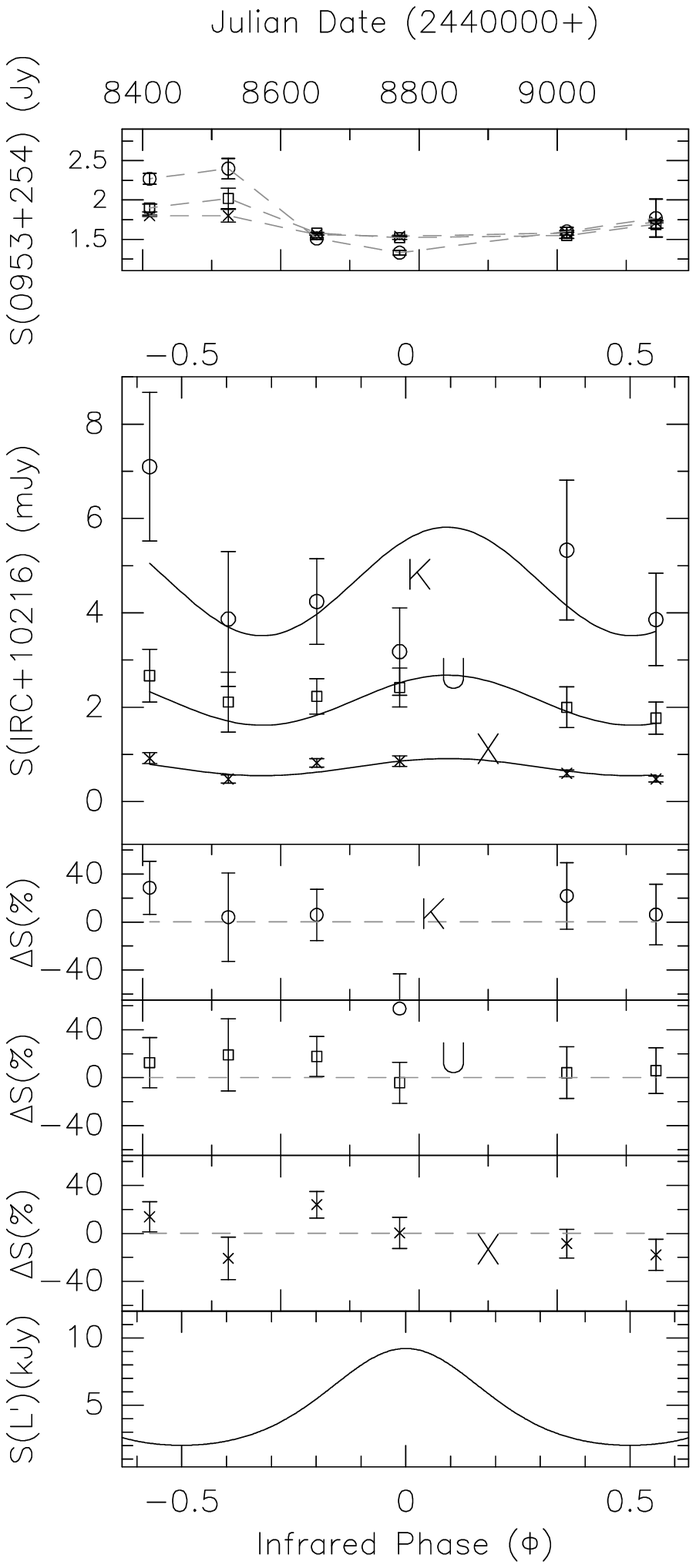}
\end{center}
\caption{\label{vari} Radio flux densities of IRC+10216 from 1991 June 1 to
1993 June 8 (\textit{second panel from top}) compared to that of the
calibrator 0953+254 (\textit{top panel}).  For both sources the X-, U-, and
K-band integrated flux densities from Gaussian fits are plotted
\textit{bottom} to \textit{top}.  Note the different flux density scales.
Error bars represent the quadratic sum of formal fitting errors and
estimates of the errors due to absolute flux density calibration
(10, 15, and 20\%\ for the X-, U-, and K-band data points).
In the IRC+10216 panel, the solid lines
represent the results of the model calculations described in the text.
The \textit{third to fifth panels from top} show K-, U- and X-band
residuals in percent--for each epoch the flux density predicted by
the model was subtracted from the measured value and divided by the
model value.
X-, U-, and K-band data points are represented as crosses, squares, and circles,
respectively. In all panels the same time range is shown.
The abscissa  gives Julian day number in the \textit{top panel} and infrared
($L'$-band) stellar phase, $\Phi$, in the others.
The date of maximum infrared light ($\Phi=0$) was calculated from the value
determined by LeBertre 1992 (JD 2447483), using his period of 649 d.
The \textit{lowermost panel} represents the $L'$-band ($3.76~\mu$m) light curve converted
to kJy units (see Fig. 2.k of \citet{LeBertre1992}).
}
\label{fig2}
\end{figure}

In Fig. \ref{vari} we show the flux densities of IRC+10216 listed in Table 2.
The error bars include contributions from random noise and systematic
calibration uncertainties.  The data show probable variability.
We fit the flux densities of IRC+10216 with a model that
includes a constant and a sinusoidally varying component.
Rather than fitting the data from each band separately,
we assume that a constant spectral index, $\alpha$, describes the difference
in flux density with observing frequency.  Specifically, the
model for the flux density, $S(\nu,t)$, at a frequency, $\nu$,
and time, $t$, is given by the following:
$$ S(\nu,t) = \bigl(S(\nu_0) + \Delta S(\nu_0) \cos{(\Delta t/P)}
             \bigr)~\bigl(\nu/\nu_0\bigr)^\alpha~~~,$$
where $S(\nu_0)$ and $\Delta S(\nu_0)$ are the constant and
amplitude of the time varying flux density at a reference frequency
$\nu_0$ (set to 10 GHz), $\Delta t$ is the time offset from maximum
($t_{\rm max}$), and $P$ is the period.

A weighted least-squares fit to the data yielded the following
results:
$S(\nu_0) = 1.02\pm0.14$~mJy,
$\Delta S(\nu_0) = 0.25\pm0.09$~mJy,
$P = 535\pm51$~d,
$t_{\rm max} = ({\rm JD} 2440000+) 8841\pm24$~d,
and $\alpha = 1.88\pm0.14$.
This fit has a reduced $\chi^2$ per degree of freedom of
1.6, and the parameter uncertainties are the formal $1\sigma$
values scaled by $\sqrt{1.6}$.  Thus, IRC+10216 appears to
display flux density variations of $\pm25$\% at a significance level
of nearly $3\sigma$.
The variability will be further discussed in the Appendix.

IRC+10216 shows strong periodic variability at infrared wavelengths
\citep{Dyck1991, LeBertre1992}.
Dyck et al. determine a period of 638 d at (infrared) $K$-band,
while Le Bertre's measurements in the $J$-, $H$-, $K$-, $L'$-, and $M$-bands yield values between
636 and 670 d.  If we assume the infrared period has a 5\% uncertainty,
then the cm-wave period of $535\pm51$~d
differs from the infrared period by about $2\sigma$.
The cm-wave maximum occurs at $8841\pm24$, whereas the
infrared maximum (extrapolated by about 1200~d) would be predicted
to occur near $8770$ with an estimated uncertainty of about $\pm50$.
Thus, the cm-wave minus infrared maxima differ by $71\pm71$~d
and are consistent within their joint uncertainty.

\subsection{Spectral index}

As discussed above, the radio emission from IRC+10216 is probably variable.
Therefore, we calculate the X--U--K-band spectral index (SI) for each
epoch and find values between 1.6 and 2.2, with uncertainties of up to 20\%.
These daily SIs are consistent with a constant value, and a variance-weighted
average of these indices is $1.92\pm 0.11$.
We also listed in Table \ref{runs} variance-weighted, time-averaged flux
densities for each band.  Using these to determine a SI yields $1.91\pm 0.11$.
The close correspondence between the daily and time averaged SIs suggests that
the spectral index is not strongly variable.

The cm-wave spectral index obtained by least-squares fitting of all of the data
simultaneously with the model described above is $1.88\pm0.14$.
This value is consistent with those observed for Mira variables
and with the value of 1.86 for a model of Miras \citet{reid1997a}.
However, the cm-wave spectral index is also consistent with a value
of 2.0 for optically-thick black-body emission.

At millimeter- and sub-millimeter wavelengths, flux densities measured
with single-dish telescopes (with $> 10''$ FWHM beams) contain contributions
from the star as well as from the extended dust envelope.
If the measurements are made with wideband bolometer detectors,
circumstellar molecular emission may also contribute $\approx30$\%\
to the total flux density \citep{Walmsley1991, Groenewegen1997}.
Whereas emission by dust dominates in the submillimeter region on large
scales (see Fig. 4), it is negligible at cm--wavelengths.  Based on the model
discussed in the Appendix, we expect that the dust envelope contributes only
$\sim$0.1 mJy to the total flux density of $\sim$2 mJy in our VLA K-band
measurements.

\begin{figure}[h]
\begin{center}
\includegraphics[width=7cm,angle=0]{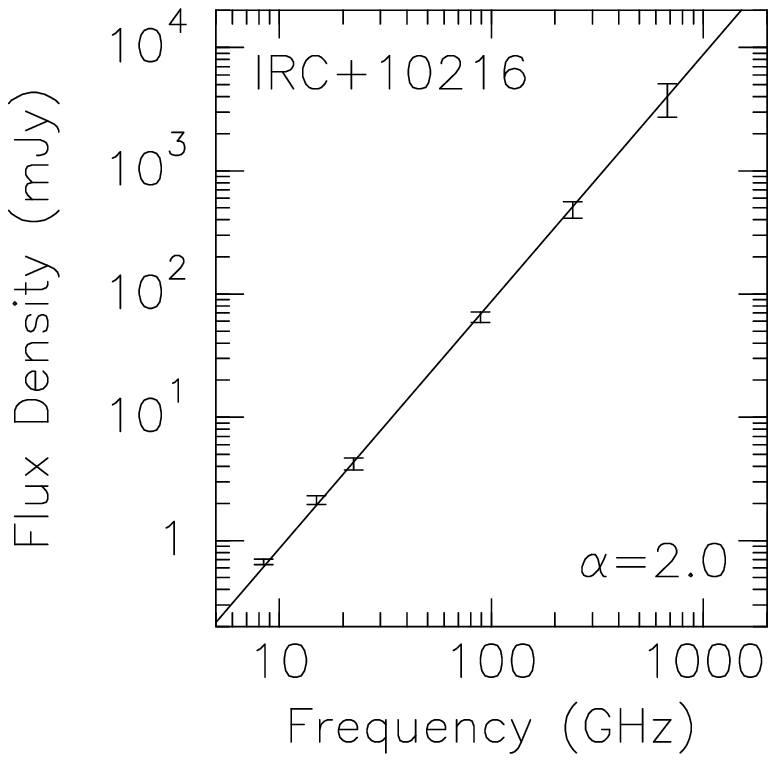}
\end{center}
\caption{\label{si} Interferometric flux densities of IRC+10216 from
cm to sub-mm wavelengths.  The X-, U-, and K-band cm-wave data
points are the weighted averages of data from six epochs (see Table 1). These
error bars are quadratic sums of formal uncertainties and
absolute calibration errors, which are assumed to be 10, 15, and 20\%\ for the
X-, U-, and K-band data, respectively.  The error bars of the mm and submm
wavelength data are discussed in the text.
The  straight line has a slope of 2.0, the blackbody value. A least squares
fit to all the shown data delivers a slope of $1.96\pm0.04$.}
\label{fig1}
\end{figure}

A few interferometric observations exist that should deliver
accurate measurements of compact emission free of dust contamination.
Also by selecting portions of the band largely free of molecular lines,
nearly pure continuum emission can be measured.
\citet{Lucas1999}, using the IRAM Plateau
de Bure Interferometer (PdBI), find a ``point source'' at 89 and 242 GHz
with flux densities of $65\pm7$ and $487\pm70$ mJy, respectively.
(Note these uncertainties are much larger than the \textit{formal}
errors given in the cited paper; R. Lucas personal communication.)
These authors also detect ``extended'' emission
with flux densities of 12 and 57\%\ of the point source values at 89
and 242 GHz, respectively, which they ascribe to dust emission from the inner
envelope.
\citet{Young2004} used the Smithsonian Sub-Millimeter Array \citep{Ho2004} to image
the 680 GHz continuum emission of IRC+10216.  They also find a ``compact, unresolved
component'' in their $\approx2''$ FWHM beam with a flux
density of $3.9\pm1.2$ Jy.
The flux densities discussed above are plotted in Fig. \ref{fig1}.
A least-squares fit to all the data (cm through
sub-millimeter) delivers an SI of $1.96 \pm 0.04$.

\subsection{\label{astrometry}Absolute position of the radio emission distribution}

During our 1993 January 26 observations, the VLA was in its most extended
configuration and the phase coherence were excellent.
Data from that epoch can be used to obtain meaningful images of
the emission distribution and to make an astrometric position
determination. To check the validity of the size information and
the position accuracy, we determined, for the K-band data,
phase and amplitude calibration of only
the first scan of 0953+254 and then applied extrapolations of these solutions to
subsequent scans of that source, which were spaced by $\approx16$ minutes. As
expected, the phase coherence degraded with elapsed time between the first and
subsequent scans. However, in maps made from the $uv$-data of
the second scan, taken 16 minutes after the first,
the source showed $<3$\% amplitude degradation, a (formal) source size
$<31$ mas and an offset of $<5$ mas in both coordinates from the nominal
position. Since IRC+10216 scans were placed between 0953+254 scans,
we conclude that some, but probably little, blurring occurs due to deficient
phase calibration. The test described above does not account for phase calibration
errors caused by the fact that 0953+254 is $> 12\deg$ away from IRC+10216. To take that
into account, we increase our value for the systematic position uncertainty
to 20~mas.

The position uncertainty may be dominated by statistical errors due to the
modest signal to noise ratio of our continuum images.  JMFIT delivered positions
for the X-, U-, and K-band images differing by maximally 20 mas from each other,
with formal errors of between 5 and 16 mas.  The total uncertainty, obtained by
quadratically adding the systematic and statistical uncertainty is 15 mas in
each coordinate.  The variance-weighted mean position and its
uncertainty is given in Table \ref{hip}.
The position derived from the data of \citet{Drake1991} is
$(\alpha,\delta)_{{\rm J}2000}$ $= 09\h47\m57\decs382, +13\deg16'43\decas61,$
for which we estimate an uncertainty of 20 mas in each
coordinate (see \S\ref{Other}).  The latter position (measured on 1987 June
2.5) is offset from the 1993 January 26 position by ($-146,-20$)~mas
in the (East,North) directions.  Taking
these offsets at face value, we derive proper motion components of IRC+10216
of ($+26\pm6$,$+4\pm6$) mas~yr$^{-1}$, respectively.  A motion of that
magnitude is plausible.  For example, Mira ($o$ Ceti), which is at
a similar distance as IRC+10216, has a larger proper motion of
($+33,-239$) mas~yr$^{-1}$ measured by Hipparcos \citep{Perryman1997}.

The proper motion derived above is consistent with the upper limit
of  $30$  mas~yr$^{-1}$ that \citet{Becklin1969} derive from a comparison
of a \textit{Palomar Sky Survey} plate taken in 1954 with a plate they
took in 1969.


\subsection{\label{morph}Morphology}
Our best quality radio data (1993 Jan 26) suggest an
elongated flux distribution (position angle, P.A., $82\pm5$ degrees; i.e.
roughly east-west direction), with major and minor axes
of $213\pm20$ and $73\pm20$ mas,
respectively, which is significantly larger than the 1987 upper limits
of 95 mas discussed in \S\ref{Other}.


While an elongated source would be tantalizing, we have greater confidence
in the 1987 than the 1993 data.
The separation between IRC+10216 and 0952+179, the phase calibrator used
by Drake et al., is only $1\decdeg7$ in right ascension and $4\decdeg4$
in declination.  In contrast, the phase calibrator used in 1993
is displaced by $2\decdeg1$ in right ascension
and a much larger distance, $-12\decdeg2$, in declination from IRC+10216.
While it is difficult to tell conclusively,
we believe that the source size observed in 1993 January
may be affected by uncorrected residual phase errors stemming from the
large angular separation of IRC+10216 and the calibrator.

\citet{Groenewegen1996} takes the elongation found by \citet{Drake1991} at
face value and argues for asphericity.
We note that on scales of a few stellar radii, IRC+10216's
morphology is highly asymmetric and changes with time.  Multi-epoch 2.2 $\mu$m
speckle interferometry between 1995 and 1998 shows an asymmetric clumpy shell
with two dominant components, A and B, whose separation grows from 191 to 265
milliarcseconds (mas) between 1995 and 1998 \citep{Osterbart2000}.  2.2 $\mu$m
Keck I aperture masking interferometry between 1997 and 1999 confirm the
observed morphology \citep{Tuthill2000}.  While both studies find motions in the
material, these are inconsistent with simple spherical expansion and a
bipolar flow.
\citet{Osterbart2000} suggest that the star itself is identical to, or located
near, component B.  On the other hand, \citep{Richichi2003}, modeling their
infrared lunar occultation data, favor an identification of the star with
component A.

Future VLA observations outlined in \S\ref{outlook}, together with
absolute infrared astrometry, may unambiguously establish the star's position
relative to the clumpy components of its envelope.
Also, future
VLA array observations at 1.3 and 0.7 cm can definitively settle the
question of IRC+10216's radio source size and morphology.



\section{\label{nature}The nature of the radio emission}

\citet{Groenewegen1997} performed model calculations of IRC+10216's infrared through radio continuum
emission. His models and assumptions under-predict our measured radio flux
densities by a factor of $\approx1.8$.
A nearer distance of $\approx100$~pc (instead of the 135 pc assumed)
would provide a better fit to the data.
For X- and U-band, his models predict the emission to arise solely
from the stellar photosphere, and at K-band the dust contribution
\textit{from the whole envelope} would add about 25\%\ to the photospheric emission.
Assuming an effective temperature of 2000 K, a photospheric diameter of 70.2 mas
is derived, which is consistent with our upper limit. As shown in the Appendix,
assuming these values for temperature and diameter, IRC+10216's spectral energy
distribution can be modeled successfully.

A basic question regarding the cm-wave emission is whether it
comes from the stellar photosphere or a somewhat larger radio photosphere.
The radio emission from Mira variables comes from a radio photosphere,
which is controlled by H$^-$ free-free opacity and is
approximately twice the size of the stellar photosphere
(defined in line-free regions of the IR spectrum) \citep{reid1997a}.
If a size (or stricter upper limit) could be measured
with the VLA, that would directly yield a brightness temperature.
A temperature below 2000 K would be expected for a
radio photosphere, whereas a temperature in
excess of 2000 would be expected for the stellar photosphere.

The amplitude of the radio variations of IRC+10216 of $\pm 25$\% is larger
than for Mira variables, where variations of $<\pm15$\%
are observed by \citet{reid1997a}.   This is in contrast to
variations of $\pm40$\% (see appendix) that would be expected from the
stellar photosphere of IRC+10216, based on IR variations of nearly 2 magnitudes.
For Miras, simple models of radio photospheres with shocks propagating
outward with speeds less than $\approx7$~\kms\ can explain
the absence of strong radio variability.   However, shocks propagating
outward with speeds of $\approx10-15$~\kms\ can
produce variation of the magnitude we find in IRC+10216
\citep{reid1997b}.

\section{\label{outlook}Conclusions and outlook}

Using the VLA, observations of IRC+10216 at 8.4, 14.9, and 22.5 GHz were made
over a three year period.  We find probable variability and a spectral
index of about 1.9, i.e. near that of optically thick blackbody emission,
but consistent with a radio photosphere as observed in Mira variables.

Future VLA A-configurations observations at Q-band (7 mm wavelength)
with a resolution of 35 mas should definitely resolve the emission
and yield a measurement of the brightness temperature.  This could
discriminate between a stellar and radio photosphere.  Such observations
may even lead to the detection of surface structure,
as observed for $\alpha$ Orionis by \citet{Lim1988}.
These observations will benefit greatly by employing the
``fast-switching'' technique.

\section*{Appendix -- Modeling IRC+10216's Spectral Energy Distribution}

The radiative transfer in the dusty circumstellar envelope around IRC+10216 has
been computed by various authors (for instance, \citet{LeBertre1987,
Groenewegen1997}), but the most ambitious and complete modeling was performed by
\citet{Menshchikov2001} who {\it a)} considered a two--dimensional configuration
to explain the non--radial structure close to the star (discovered
at 2.2 $\mu$m wavelength speckle interferometry by \citet{Osterbart2000}),
{\it b)} took into account details in the composition of the dust by adding,
to the major component of amorphous carbon, admixtures of silicon carbide
and magnesium sulfide grains to reproduce broad spectral features seen by the
Infrared Space Satellite (ISO) and from the ground (at 11 and $\sim$30$\mu$m)
and, {\it c)},
included the phase in the stellar cycle.  The bolometric luminosity of the star
oscillates in their model by a factor of 2.5, although the average (infrared) $K$-band
variations of $\sim$2\,mag (Dyck et al.~1991) would suggest a greater ratio of
maximum to minimum luminosity, $L_{\rm max} / L_{\rm min} \simeq 6$.

We add to the existing radiative transfer calculations another one, on a much
more modest level, which has the goal of clarifying the relative contributions
of the star and its dusty envelope to the long wavelength emission.  In
light of the huge amount of observational data, simplicity has, despite its
obvious shortcomings, the advantage of transparency and of not being entangled
by occasionally secondary details.  The numerical code is based on a standard
ray tracing method assuming radial symmetry.
It can handle an arbitrary number of
grain types, includes scattering and is described in section 13.2 of \citet{Kruegel2003}.

Our model parameters are as follows: The star has at maximum a luminosity $L =
1.1\times 10^4$ L$_\odot$ and emits as a blackbody of temperature
$T_*=2000$\,K.  The density distribution of the dust is radially symmetric with
$\rho(r) \propto r^{-2}$.  The outer radius of the circumstellar envelope equals
$3\times 10^{17}$~cm, corresponding to a source diameter of 6$'$ at a
distance of 110~pc, the inner radius is $3\times 10^{14}$ cm, or about
five stellar radii.  There the dust temperature is around 1000 K, which is close
to the expected condensation temperature.  The optical depth in the $V$-band
toward the star amounts to $A_{\rm V}= 48$ mag.  Neglecting minor mineralogical
components, the dust consists only of spherical amorphous carbon grains with a
$n(a) \propto a^{-3.5}$ size distribution and lower and upper limits
$a_-=80$\AA, $a_+=640$\AA, respectively, and optical constants as compiled by
\citet{Zubko1996} for their type BE.  At 1mm wavelength, the mass absorption
  coefficient, $K_\nu$, is about 0.5 cm$^2$ per g of dust with a $\nu^{1.6}$
  frequency dependence.  As the grains are much smaller than the wavelength,
  their size distribution does not affect the value of $K_\nu$ at FIR or longer
  wavelengths.  Being a radiative transfer model for a dusty medium, the
  dust--to--gas ratio does not enter our calculations.

\begin{figure}[h]
\begin{center}
\includegraphics[width=8cm,angle=0]{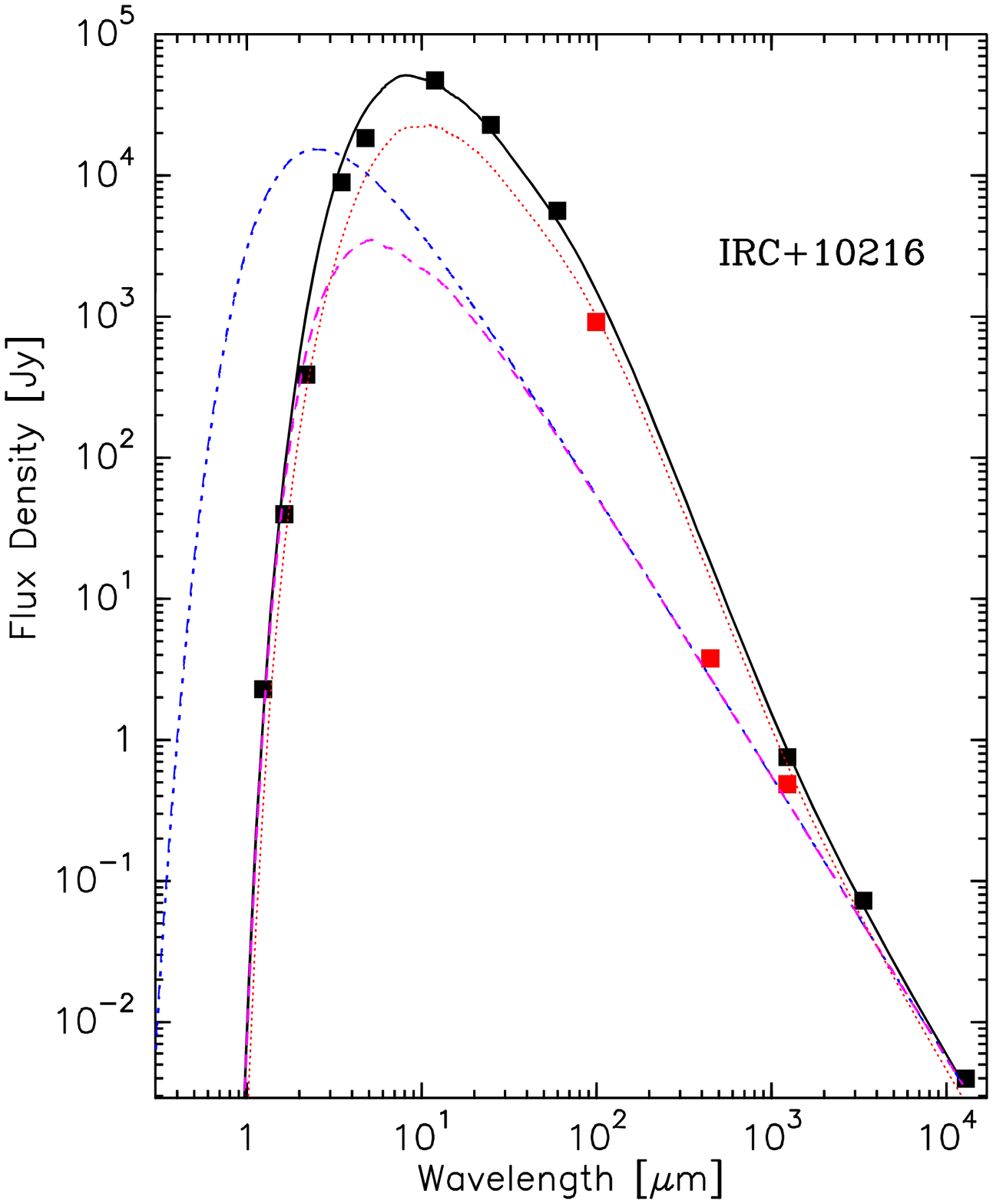}
\end{center}
\caption{\label{sed} Spectral energy distribution for IRC+10216.
{\it Data} (squares): $J$- to $M$- band
from Le Bertre (1987); 12 to 100$\mu$m from IRAS; 450$\mu$m from Young et
al.~(2004) for 2$''$ beam; 1.24mm fluxes from Lucas \& Guelin (1999), lower
point for 1.$''$6 beam, upper point for total envelope; 88\% of the total 3mm
flux (same authors) come from the inner 2.$''$4; 1.3 cm point, this paper.
{\it Models:} The solid curve is the  model described in the appendix for
maximum luminosity, the
dotted lines are for minimum light, when the stellar luminosity is 2.5 times
weaker and the surface temperature is 1700 K.
The dash--dots, which overshoot the
near IR data points, are flux densities from the star alone.
The dashed line gives
only the stellar flux, after extinction by the envelope of $A_{\rm V}=
48$~mag.  }
\end{figure}

Figure \ref{sed} displays model results together with observational data
appropriate near maximum luminosity.  The solid curve is our fit near
maximum luminosity within a 6$'$ beam.  It was obtained by adjusting the total
optical depth, the inner radius, and the grain sizes, the other above mentioned
parameters were kept fixed.  Overall, the fit is satisfactory.  The squares
which lie significantly below the model refer to smaller angular sizes (see
the caption of Fig.~\ref{sed}).

\begin{figure}[h]
\begin{center}
\includegraphics[width=9cm,angle=0]{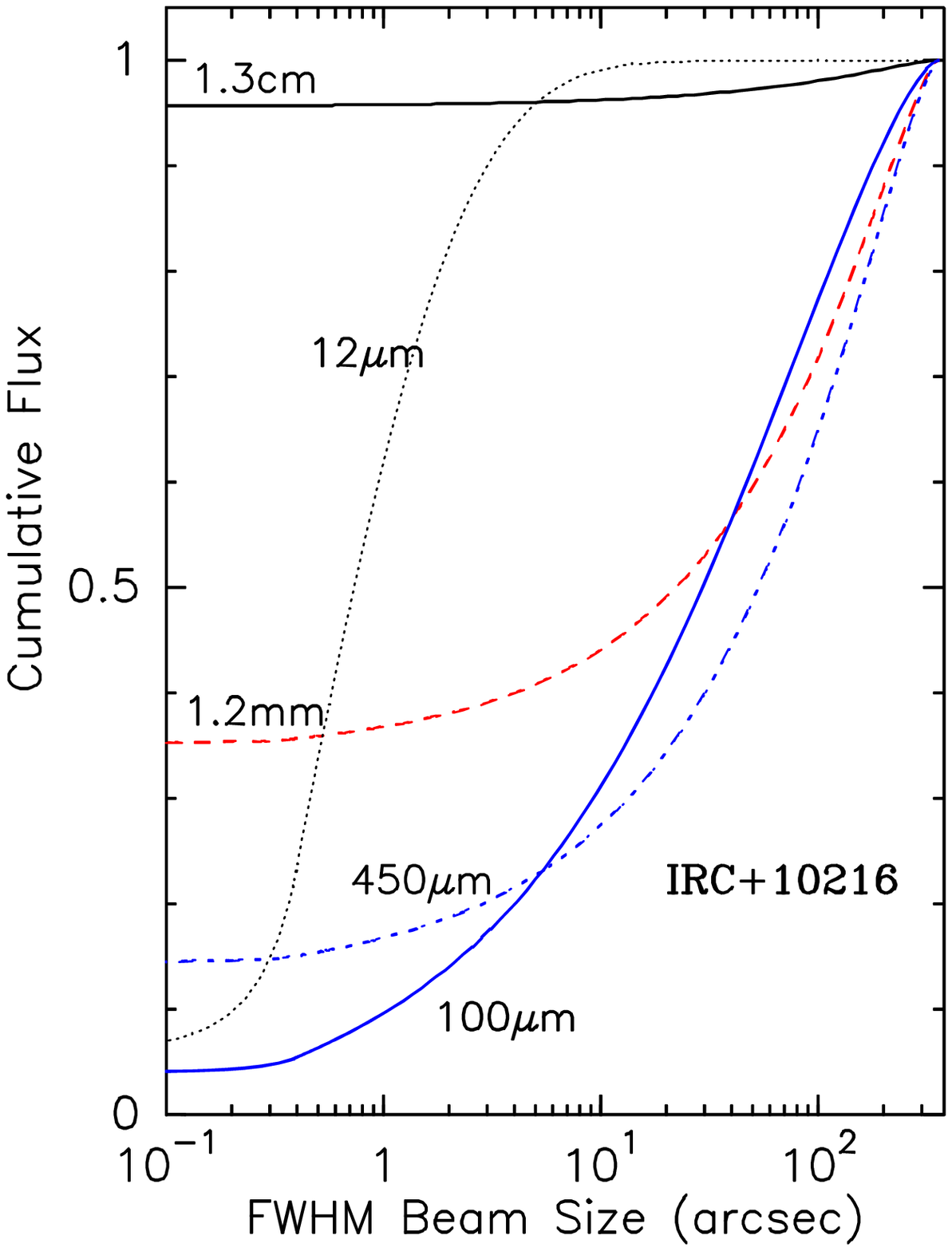}
\end{center}
\caption{\label{normalized_flux}
Normalized model flux density as a function of aperture at five wavelengths.
The star has an assumed angular diameter of 0.''075; at a comparable
resolution essentially all the emission comes from the star.
Substantial dust emission is indicated for beam sizes $>1''$ at IR wavelengths.
}
\end{figure}

Figure \ref{normalized_flux} shows how the flux increases with beam size.  As
the curves are flat in the very left part of the figure, the emission from
diameters less than 0.$''$2 is due to the star, not to dust, despite the central
dust density peak.  One can read from the figure that the fraction of
the total flux contributed by the stellar photosphere
increases from $\sim$4\% at 100$\mu$m to $\sim$96\% at 1.3 cm.

One can qualitatively understand why the brightness variations over a
pulsational period are considerably smaller in the radio regime
(Fig.\ref{vari}) than at near infrared wavelengths.
The monochromatic flux density which we would
receive without foreground extinction, $F_\nu$,
from a star with a blackbody photosphere
of temperature $T$ and luminosity $L$ is
$$ F_\nu = {L B_\nu(T) \over 4\sigma D^2 T^4} $$
where $\sigma$ denotes the Stefan-Boltzmann constant and $D$ the distance.  The
equation becomes obvious by integrating over frequency which yields $4\pi D^2 F
= L$, with $F=\int F_\nu d\nu$.  In the Rayleigh--Jeans limit, at radio
wavelengths, $B_\nu(T) \propto T$ and therefore $F_\nu \propto L / T^3$.

How do $L$ and $T$ vary during a cycle?  Reid \&\ Menten (1997a) in their Eq.(1),
which is based on observations by \citet{pettit1933},
approximate the stellar radius and temperature of a Mira variable as a function
of phase $\phi$ by $T(\phi) = 2300 + 300 \cos\phi$ K and $R(\phi)= 2 + 0.4
\sin\phi$ AU.  Oscillations in luminosity, $L(\phi)$, follow
because $L=4\pi R^2\sigma T^4$.  According to these formulae, temperature and
luminosity are roughly in phase, and therefore the observed radio flux $F_\nu
\propto L / T^3$ changes less than $L$ itself.
The radio $F_\nu$ varies from maximum to minimum by a
factor of 2 (corresponding to excursions from the mean by $\pm40$\%)
and $L$ varies by a factor 3.76.
We note that at infrared wavelengths, assuming a
constant optical depth over the cycle and no dust emission, the monochromatic
flux mimics $L$ almost closely.
The dotted curve in
Fig.\ref{sed} is our model for IRC+10216 at minimum (2.5 times less
luminous, $T=1700$ K).  At minimum, the 1.3 cm radio brightness is
25\% lower and the $J$- and $H$-band flux densities are 2.5 times lower
compared to maximum light.


The radio variations around the mean suggested by Fig.\ref{vari} are
smaller than $\pm40$\%.
This suggests that the cm-wave emission may not come directly from the
stellar photosphere.  However, other
possible reasons for this discrepancy might be a
smaller variability in the bolometric luminosity (by a factor 2.5, as
Men'shchikov et al.~(2001) assume, and not by 3.76) or inaccuracies in the
simple expressions for $R(\phi)$ and $T(\phi)$.

\acknowledgements{We would like to thank the referee for comments and, in particular,
Malcolm Walmsley for taking the editor's
job very seriously, which led to a significant improvement of the paper.
This research has made use of the SIMBAD database,
operated at CDS, Strasbourg, France.}

\bibliographystyle{aa}
\bibliography{4463bib}

\clearpage

\end{document}